# A Stepped Oxide Hetero-Material Gate Trench Power MOSFET for Improved Performance

Raghvendra S. Saxena and M. Jagadesh Kumar, *Senior Member, IEEE*

**Abstract:** In this work, we propose a new Stepped Oxide Hetero-Material Trench (SOHMT) power MOSFET with three sections in the trench gate (an $N^+$ poly gate sandwiched between two $P^+$ poly gates) and having different gate oxide thicknesses (increasing from source side to drain side). The different gate oxide thickness serves the purpose of simultaneously achieving (i) a good gate control on the channel charge and (ii) a lesser gate to drain capacitance. As a result, we obtain higher transconductance as well as reduced switching delays, making the proposed device suitable for both RF amplification and high speed switching applications. In addition, the sandwiched gate with different work function gate materials modifies the electric field profile in the channel resulting in an improved breakdown voltage. Using two-dimensional simulations, we have shown that the proposed device structure exhibits about 32% enhancement in breakdown voltage, 25% reduction in switching delays, 20% enhancement in peak transconductance and 10% reduction in figure of merit (product of ON-resistance and gate charge) as compared to the conventional trench gate MOSFET.

Final Version of Accepted Manuscript MS# 6968R

## I. INTRODUCTION

Trench gate MOSFETs are popular in various medium to low voltage power applications like automotive electronics, microprocessor power supplies, DC-DC converters and control switching due to their reduced conduction power losses and forward voltage drop [1-13]. However, simultaneously achieving the other important features of a power device e.g. high breakdown voltage, high transconductance and high switching speed, is difficult as any improvement in one performance parameter degrades the others. For example, in a standard trench structure, increasing the gate oxide thickness reduces gate-to-drain capacitance and, therefore, improves the frequency response but at the same time, it reduces the gate control on the channel charge and lowers the transconductance. Similarly, any attempt to increase the breakdown voltage causes a drastic increase in the ON-resistance [14, 15]. Structural modifications are often required to overcome these limitations and optimize the device performance [1-10].

In this paper, we propose a novel trench gate structure, termed here as SOHMT (Stepped Oxide Hetero Material Trench) power MOSFET in which the trench gate is composed of three sections of different gate oxide thicknesses (lower on the source side and higher on the drain side) and different gate electrode materials realized by sandwiching a low work function gate material ($N^+$ poly) between two higher work function gate materials ($P^+$ poly). By performing two-dimensional numerical simulations in ATLAS device simulator [16], we demonstrate that the use of a sandwiched gate with different work functions causes additional peaks in the channel electric field resulting in an increased breakdown voltage without adversely affecting the ON-resistance of the device when compared to the conventional trench gate power MOSFET. We also show that the smaller gate oxide thickness on source side improves the gate control of the channel charge (increasing the transconductance) and the larger gate oxide thickness on the drain side lowers the gate-to-drain capacitance (improving the switching speed). We demonstrate that in the proposed SOHMT structure, the figure of merit (FOM) defined as the product of ON-resistance and gate charge [17, 18], may be reduced about 10% along with the improvements in other performance parameters (e.g. breakdown voltage, switching delays and peak transconductance) as compared to the conventional trench gate power MOSFET that has single gate material and uniform gate oxide thickness.

## II. DEVICE STRUCTURE AND PROPOSED FABRICATION PROCEDURE

Fig. 1(a) shows the schematic cross-section of the proposed SOHMT MOSFET containing three sections of the trench (upper, middle and bottom). The bottom section has the largest gate oxide thickness of 75 nm with $P^+$ poly as the gate material. The middle section is made of $N^+$ poly gate material with 50 nm gate oxide thickness. The upper section has the smallest gate oxide thickness of 25 nm with $P^+$ gate material. All the three gate materials are connected by a metal. The equivalent conventional device used for the comparison with SOHMT device is shown in Fig. 1(b) that has single material gate of $N^+$ poly and uniform gate oxide of 50 nm thickness with all the other physical lengths and doping profile same as that of the SOHMT device.

The fabrication process of the SOHMT structure is shown in Fig. 2. First, using the conventional processing sequence of a trench MOSFET [11-13], we create the structure as shown in Fig. 2(a). It contains the $N^+$ ($N_D = 1 \times 10^{19}$ cm$^{-3}$) substrate as drain and 0.1 μm thick $N^+$ ($N_D = 1 \times 10^{19}$ cm$^{-3}$) source on top surface, a 2.5 μm thick N-type ($N_D = 1 \times 10^{16}$ cm$^{-3}$) drift region and a 0.5 μm thick P-type ($N_A = 5 \times 10^{17}$ cm$^{-3}$) body region. A 75 nm thick gate oxide layer is grown inside the trench whose width and depth are 1.0 μm and 1.3 μm, respectively. A blanket deposition of $P^+$ poly silicon and CMP will result in a structure shown in Fig. 2(b). A selective etching of poly-silicon followed by oxide etching will form the 0.3 μm thick $P^+$ poly bottom gate electrode as shown in Fig. 2(c). Again, using similar steps (gate oxide formation, poly deposition, CMP and etching) we realize the middle and bottom gate sections as shown in Fig. 2(d) that contains 50 nm thick middle gate oxide and about 0.6 μm thick $N^+$ poly middle gate electrode and 25 nm thick top gate oxide and $P^+$ poly top gate electrode up to the brim of the trench. Now, we make a contact hole of 0.5 μm width and 1.1 μm depth in the middle of the trench as shown in Fig. 2(e), which is then filled with a metal using standard metallization process to connect all the three gate electrodes to result in the final structure of the SOHMT device, shown in Fig. 2(f). The conventional device studied here for comparison also has same doping and geometrical parameters except the gate structure as shown in Fig 1(b). It can be fabricated using process steps up to Fig. 2(b) with 50 nm gate oxide and $N^+$ poly deposition followed by metallization process.

The additional steps required for the SOHMT device are: deposition, etching and CMP process steps and also one additional mask for contact hole opening as compared to the conventional device. The improvements achieved in the performance parameters (as discussed in subsequent sections) are at the cost of the above additional process steps which are not very complex.

### III. SIMULATION RESULTS AND DISCUSSION

We have created the SOHMT device and the conventional device in ATLAS device simulator and performed numerical simulations for their current-voltage characteristics, breakdown performance and switching speed. Due to the difference in the gate oxide thickness and in the work function of the gate materials, the energy bands are modified in the SOHMT MOSFET. The conduction band energy along the trench side wall is shown in Fig. 3 in contrast with the conventional device for $V_{DS} = 1.0$ V and $V_{GS} - V_T = 1.0$ V. It may be noticed that the potential barrier for the electrons moving from the source into the channel reduces in the SOHMT device.

Furthermore, the abrupt changes in the surface potential at the boundaries (boundary between top and middle gate, and middle and bottom gate) modify the channel electric field as shown in Fig 3 resulting in the additional peaks in the electric field profile of the SOHMT device in comparison to the conventional device. The changes in the energy bands and the electric field profile lead to significant improvements in the device characteristics as discussed below.

A. *Current Voltage Characteristics*

The output characteristics ($I_{DS}$-$V_{DS}$) for different gate over-drive voltages are shown in Fig. 4(a), depicting the higher drive current in the SOHMT device as compared to the conventional device for all the bias conditions. This improvement occurs because of the smaller potential barrier for the electrons while moving in the channel towards the drain as shown in the conduction band profile in Fig. 3. Due to the higher drive currents the ON-resistance of the device decreases. On an average 19.2% improvement in specific ON-resistance has been observed in simulation results of SOHMT device as compared to conventional device as shown in Fig. 4(b). Furthermore, the SOHMT device also shows higher peak transconductance ($g_m$). This occurs not only due to a better gate control of the inversion charge but also due to the electric field peak in the p-body region that causes efficient carrier transport from the source to the channel [10]. As a result, the peak transconductance of the SOHMT device improves by approximately 20% as compared to the conventional device as shown in Fig. 4(c).

B. *Gate Charging Transient*

The total gate capacitance ($C_g$) plays an important role in deciding the maximum switching speed as it decides the time required to turn ON and OFF the transistor. The $C_g$ is constituted of the gate-to-source capacitance ($C_{GS}$) and the gate-to-drain capacitance ($C_{GD}$). The capacitive coupling of the gate with the source, i.e., $C_{GS}$ should be high to get high transconductance as it is indicative of the gate control of channel charge. On the other hand, the capacitive coupling of the gate with the drain, i.e., $C_{GD}$ should be small for improving switching speed as $C_{GD}$ works as miller capacitance [17, 18]. Both these requirements can not be met using the conventional device structure. In the proposed SOHMT device, the gate oxide thickness at the source side is smaller and at the drain side, it is larger than that of the conventional device. Therefore, the SOHMT device is expected to show higher $C_{GS}$ and lower $C_{GD}$ as compared to the conventional device. To study these effects we have performed the gate charging transient analysis [17-20] using mixed mode simulations in ATLAS device simulator. We have connected a 10 μA constant current source at the gate of both the devices and studied the resulting changes in gate to source voltages ($V_{GS}$) as shown in Fig. 5. The circuit configuration used in our simulation is also shown in the inset of Fig. 5. The device width has been kept at 10,000 μm for both the SOHMT and the conventional devices in this simulation. The initial part of the curve (till the slope changes) represents the charging time of $C_{GS}$. The subsequent smaller slope region indicates the time required to charge the $C_{GD}$. The charging time multiplied by the constant current forced into the gate terminal gives the amount of charge being injected into the gate. The top x-axis in Fig. 5 shows the gate charge. These curves show that $Q_{GS}$ = 565 pC/mm$^2$ and $Q_{GD}$ = 1905 pC/mm$^2$ in the SOHMT device as compared to $Q_{GS}$ = 255 pC/mm$^2$ and $Q_{GD}$ = 2205 pC/mm$^2$ in the conventional device, which amounts to about 121% enhancement in $Q_{GS}$ and 13% reduction in $Q_{GD}$ in the SOHMT device as compared to the conventional device. The improvement in the total gate charge at $V_{GS}$ = 5.0 is calculated to be 8% in the SOHMT device as compared with the conventional device. We have seen in the previous sub-section that the specific ON-resistance is reduced in the SOHMT device as compared to the conventional device. Thus, the values of $R_{ON}*Q_G$ and $R_{ON}*Q_{GD}$ also reduce by 9.6% and 15.3% respectively in the SOHMT device as compared to the conventional device.

### C. Drain Breakdown Voltage

At higher drain voltages, the electric field peak near the bottom end of the trench (i.e. the main peak) becomes much larger than the other peaks and is responsible for the breakdown in both the conventional and the SOHMT devices. The additional electric field peaks in the SOHMT device result in a reduction of the main peak as compared to the conventional device [10]. Fig. 6 shows the electric field profile of both the SOHMT and the conventional devices for a drain voltage of 130 V. It may be seen that due to the additional electric field peaks, the main peak reduces in the SOHMT device as the total area under the electric field curve has to be the same in both the cases. We find that the reduction in the main electric field is approximately $1.29 \times 10^5$ V/cm resulting in a 32% improvement in the breakdown voltage of the SOHMT device as compared to the conventional device, as shown in the inset of Fig. 6. Here, the drain voltage at which drain current becomes 50 fA/μm is defined as the breakdown voltage.

### D. Switching Speed

The switching speed is expected to improve in SOHMT device due to its smaller $Q_{GD}$ as compared to the conventional device. The switching speed improvement in the SOHMT device has been calculated using an inverter configuration as shown in Fig. 7 using mixed-mode simulation of ATLAS device simulator. The device width has been kept 10 μm for both the SOHMT and the conventional devices. It may be seen that the 50% time delay reduces from 40 ps in the conventional device to 32 ps in the SOHMT device leading to a 25% reduction in the switching delay.

The simulation results of conventional device show the performance parameters achievable using conventional device structure and those of SOHMT device show the improvement over conventional device. These improvements are summarized in Fig. 8 showing 9.6% reduction in the FOM, 19.2% reduction in ON-resistance (averaged over 0 to 5 V of gate overdrive voltage), 20.2% improvement in peak transconductance, 32.2% improvement in breakdown voltage and 25% reduction in switching delays. These improvements are achieved at the cost of some additional processing steps such as deposition, etching, CMP and contact hole opening which are not very critical.

## IV. CONCLUSIONS

We have demonstrated that in a trench gate power MOSFET, different gate oxide thicknesses may be created by sectioning the trench gate into three parts which also allows us to use different work function gate materials in these sections resulting in a novel SOHMT device structure. Using two-dimensional numerical simulations, we have shown that the proposed SOHMT device exhibits lower FOM (both $R_{ON}.Q_G$ and $R_{ON}.Q_{GD}$), higher breakdown voltage, higher transconductance, lower ON-resistance and higher switching speed compared to the conventional device. Our results indicate that it is feasible to improve all the performance parameters of a trench gate power MOSFET using SOHMT structure at the cost of few additional process steps which are not complex.

**Acknowledgement:** This work was supported in part by the International Business Machines (IBM) Faculty Award.

Raghavendra S. Saxena and M. Jagadesh Kumar "A Stepped Oxide Hetero-Material Gate Trench Power MOSFET for Improved Performance," *IEEE Trans. on Electron Devices,* Vol.56, pp.1355-1359, June **2009.**

**References**

[1] A. Narazaki, J. Maruyama, T. Kayumi, H. Hamachi, J. Moritani and S. Hine, "A 0.35um Trench Gate MOSFET with an ultra low on state resistance and a high destruction immunity during the inductive switching", in *Proc. of ISPSD-2OOO,* May 22-25, Toulouse, France, pp. 377-380, 2000.

[2] X. Yang, Y. C. Liang, G. S. Samudra and Y. Liu, "Tunable Oxide-Bypassed Trench Gate MOSFET: Breaking the Ideal Superjunction MOSFET Performance Line at Equal Column Width", *IEEE Trans. Electron Devices,* vol. 24, No. 11, pp. 704-706, Nov 2003.

[3] S. Ono, Y. Kawaguchi and A. Nakagawa, "30V New Fine Trench MOSFET with Ultra Low On-Resistance", in *Proc. of ISPSD-2OO3,* Apr 14-17, Cambridge, UK, pp. 28-31, 2003.

[4] M. Danvish, C. Yue, K. H. Lui, F. Giles, B. Chan, K. Chen, D. Pattanayak, Q. Chen, K. Terrill and K. Owyang, "A New Power W-Gated Trench MOSFET (WMOSFET) with High Switching Performance", in *Proc. of ISPSD-2OO3,* Apr 14-17, Cambridge, UK, pp. 24-27, 2003.

[5] J. H. Hong, S. K. Chung and Y. I. Choi, "Optimum design for minimum on-resistance of low voltage trench power MOSFET", *Microelectronics Journal*, vol. 35, No. 3, pp. 287-289, Mar 2004.

[6] M. H. Juang, W. C. Chueh and S. L. Jang, "The formation of trench-gate power MOSFETs with a SiGe channel region", *Semicond. Sci. Technol.*, vol 21, No. 6, pp. 799–802, May 2006.

[7] I. Cort´es, P. F. Mart´ınez, D. Flores, S. Hidalgo and J. Rebollo, "The thin SOI TGLDMOS transistor: a suitable power structure for low voltage applications", *Semicond. Sci. Technol.*, vol. 22, No. 10, pp. 1183-1188, Oct 2007.

Raghavendra S. Saxena and M. Jagadesh Kumar "A Stepped Oxide Hetero-Material Gate Trench Power MOSFET for Improved Performance," *IEEE Trans. on Electron Devices,* Vol.56, pp.1355-1359, June **2009.**

[8] R. S. Saxena and M. J. Kumar, "A novel dual gate strained-silicon channel trench power MOSFET for improved performance," in *Proc. NSTI Nano-technol. Conf. Trade Show*, Boston, MA, Jun 1–5, 2008.

[9] R. S. Saxena and M. J. Kumar, "A New Strained-Silicon Channel Trench-Gate Power MOSFET: Design and Analysis", *IEEE Trans. Electron Devices*, vol. 55, No. 11, pp. 3299-3304, Nov 2008.

[10] R. S. Saxena and M. J. Kumar, "Dual Material Gate Technique for Enhanced Transconductance and Breakdown Voltage of Trench Power MOSFETs", *IEEE Trans. Electron Devices*, vol. 56, No. 3, pp.517-522, March 2009.

[11] K. S. Nam, J. W. Lee, S. G. Kim, T. M. Roh, H. S. Park, J. G. Koo and K. I. Cho, "A Novel Simplified Process for Fabricating a Very High Density P-Channel Trench Gate Power MOSFET", *IEEE Electron Device Letters,* vol. 21, No. 7, pp. 365-367, Jul 2000.

[12] J. Kim, T. M. Roh, S. G. Kim, Y. Park, Y. S. Yang, D. W. Lee, J. G. Koo, K. I. Cho and Y. Kang, "A Novel Process for Fabricating High Density Trench MOSFETs for DC-DC Converters", *ETRI Journal,* vol. 24, No. 5, pp. 333-340, Oct 2002.

[13] M. H. Juang, W. T. Chen, C. I. Ou-Yang, S. L. Jang, M.J. Lin, H. C. Cheng, "Fabrication of trench-gate power MOSFETs by using a dual doped body region", *Solid-State Electronics*, vol. 48, No. 7, pp. 1079-1085, July 2004.

[14] B. J. Baliga, "An Overview of Smart Power Technology", *IEEE Trans. Electron Devices,* vol. 38, No. 7, pp. 1568-1575, Jul 1991.

[15] R. P. Zingg, "On the Specific On-Resistance of High-Voltage and Power Devices", *IEEE Trans. Electron Devices,* vol. 51, No. 3, pp. 492-499, Mar 2004.

[16] *Atlas User's Manual: Device Simulation Software*, Silvaco Int., Santa Clara, CA.

[17] R. J. E. Hueting, E. A. Hijzen, A. W. Ludikhuize, and M. A. A. in 't Zandt, "Switching performance of low-voltage n-channel trench MOSFETs,", in *Proc. ISPSD*, pp. 177–180, 2002.

[18] R. J. E. Hueting, E. A. Hijzen, A. Heringa, A. W. Ludikhuize and M. A. A. in't Zandt, "Gate-Drain Charge Analysis for Switching in Power Trench MOSFETs", *IEEE Trans. Electron Devices*, vol. 51, No. 8, pp. 1323-1330, Aug 2004.

[19] M. Alwan, B. Beydoun, K. Ketata, and M. Zoaeter, "Gate charge behaviors in N-channel power VDMOSFETs during HEF and PBT stresses", *Microelectronics Reliability,* vol. 47, pp. 1406–1410, Sep-Nov 2007.

[20] L. Thé´olier, K. Isoird, H. Tranduc, F. Morancho, J. Roig, Y. Weber, E. N. Stefanov, and J. M. Reyne`s, "Switching performance of 65 V vertical N-channel FLYMOSFETs", *Microelectronics Journal*, vol. 39, pp. 914–921, June 2008.

FIGURE CAPTIONS

Fig. 1: Schematic cross-sectional view of (a) The SOHMT device and (b) The equivalent conventional device.

Fig. 2: Process steps to fabricate SOHMT device.

Fig. 3: Conduction band energy and channel electric field along the trench sidewall in the channel of SOHMT and conventional devices for $V_{DS} = 1.0$ V and $V_{GS}-V_T = 1.0$ V.

Fig. 4: Characteristics of SOHMT device as compared to the conventional device: (a) output I-V characteristics for different gate over drive voltages $V_{GS}-V_T$, (b) on state resistance as function of $V_{GS} - V_T$, and (c) transconductance as function of gate voltage for different $V_{DS}$.

Fig. 5: Gate charging transient curves for SOHMT and conventional device for 10μA gate charging current.

Fig. 6: Electric field profile at $V_{DS} = 130$ V and I-V characteristics showing the breakdown condition.

Fig. 7: Switching simulation results for both SOHMT and convention devices in an inverter configuration.

Fig. 8: The bar chart indicating the percentage improvement in various parameters of the SOHMT device as compared with conventional device.

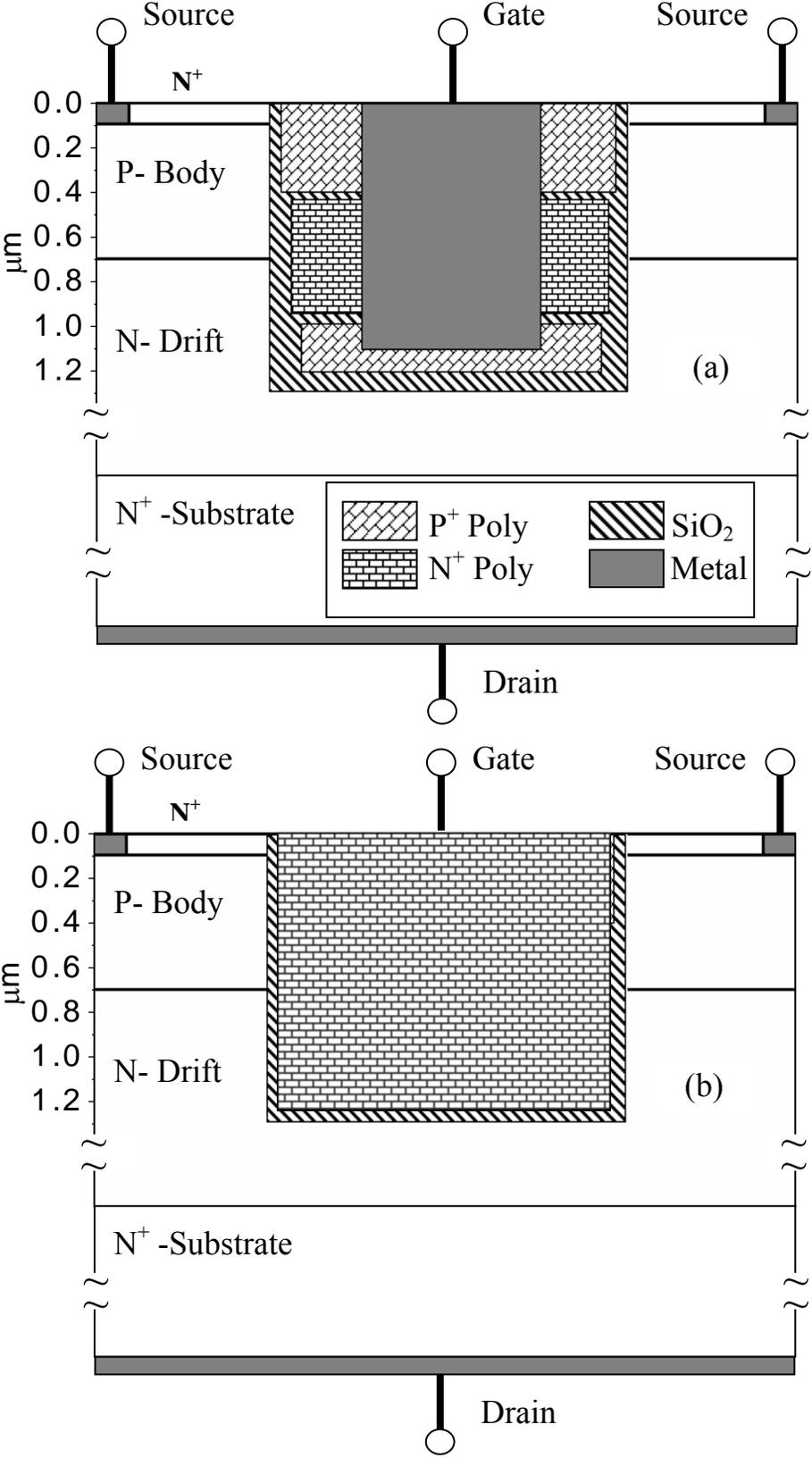

Fig. 1

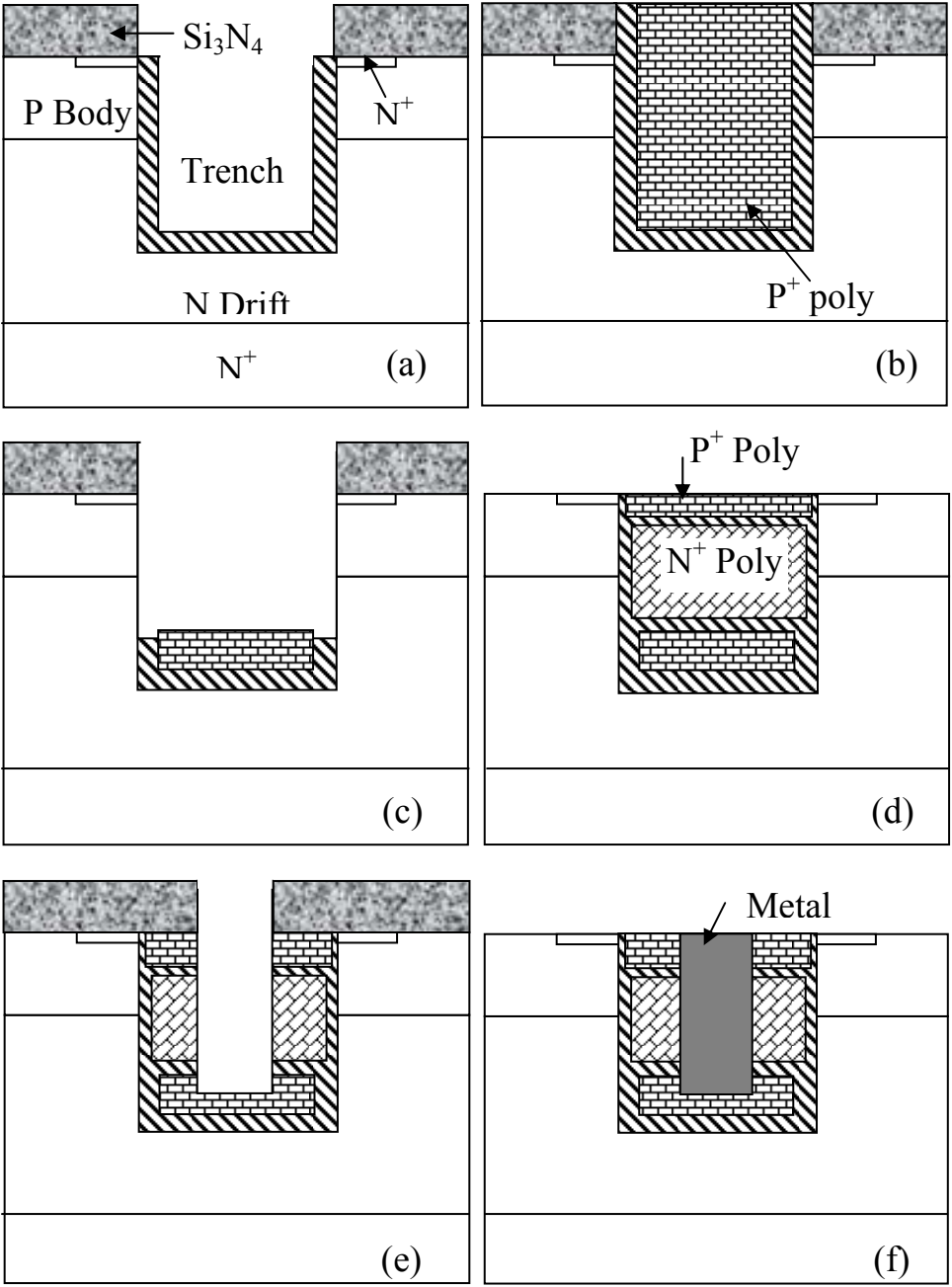

Fig. 2

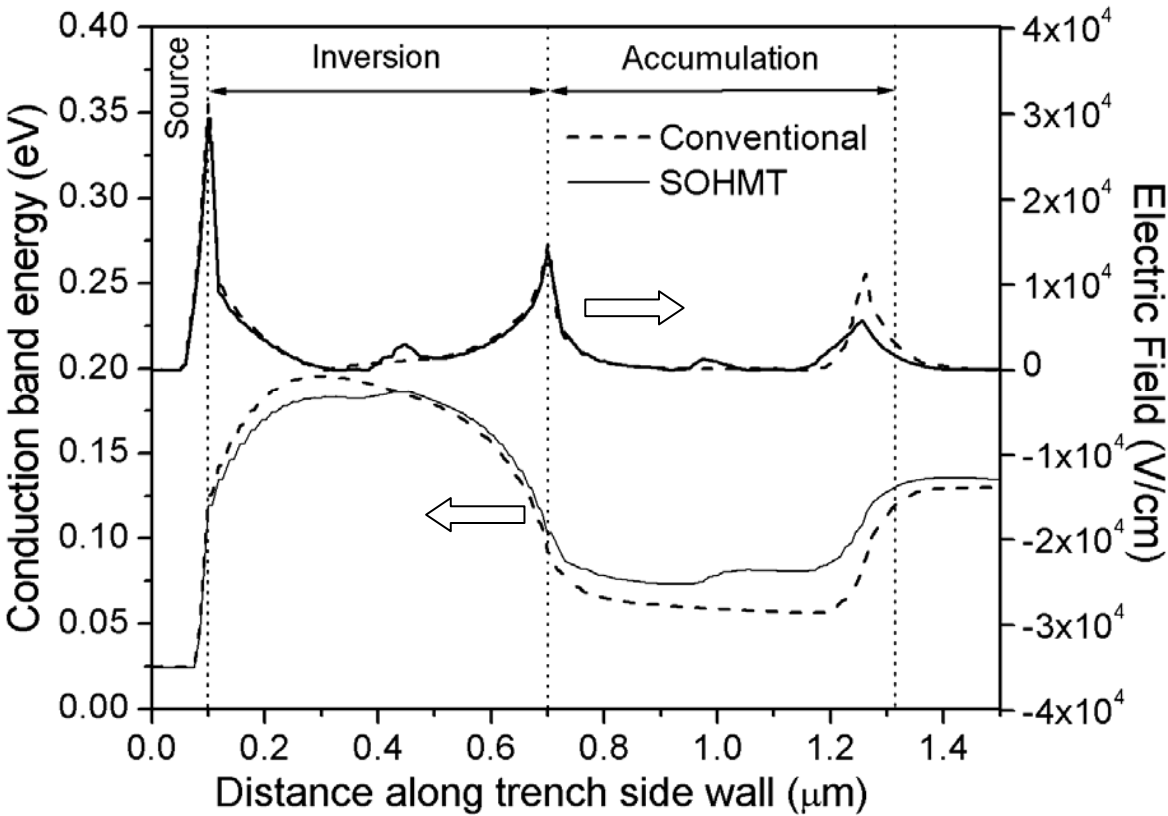

Fig. 3

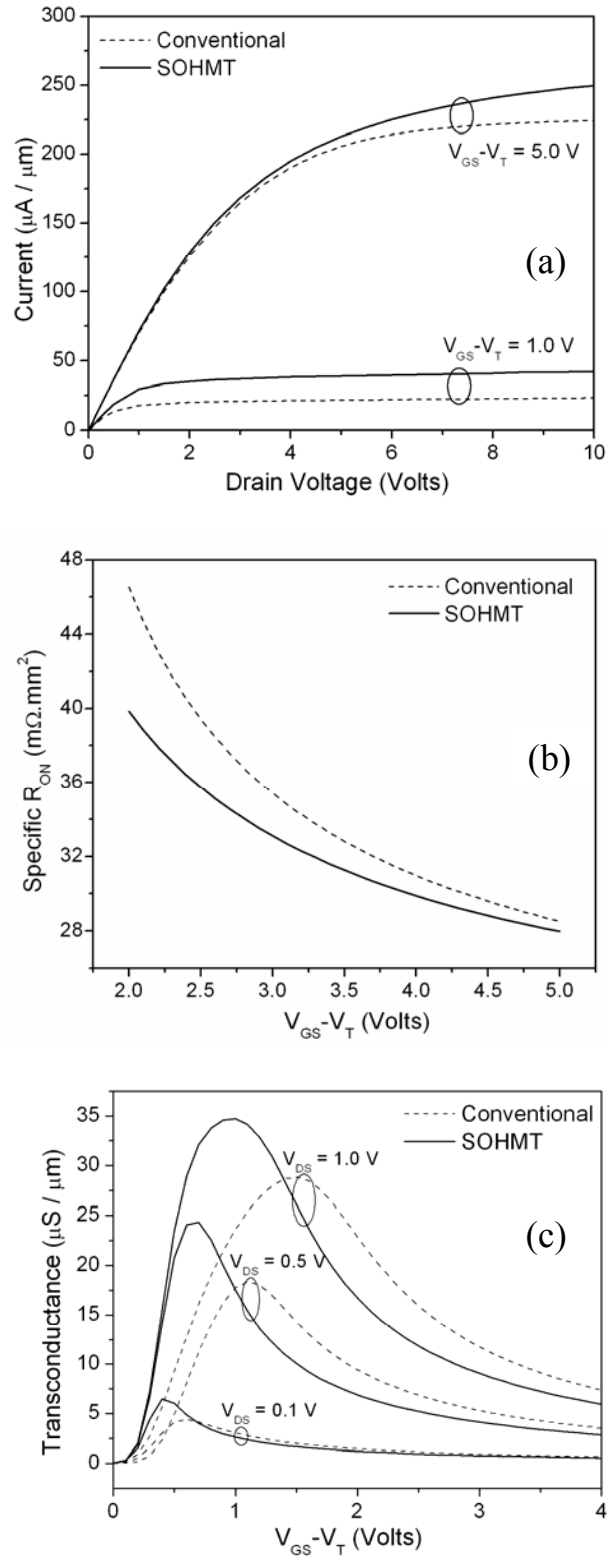

Fig. 4

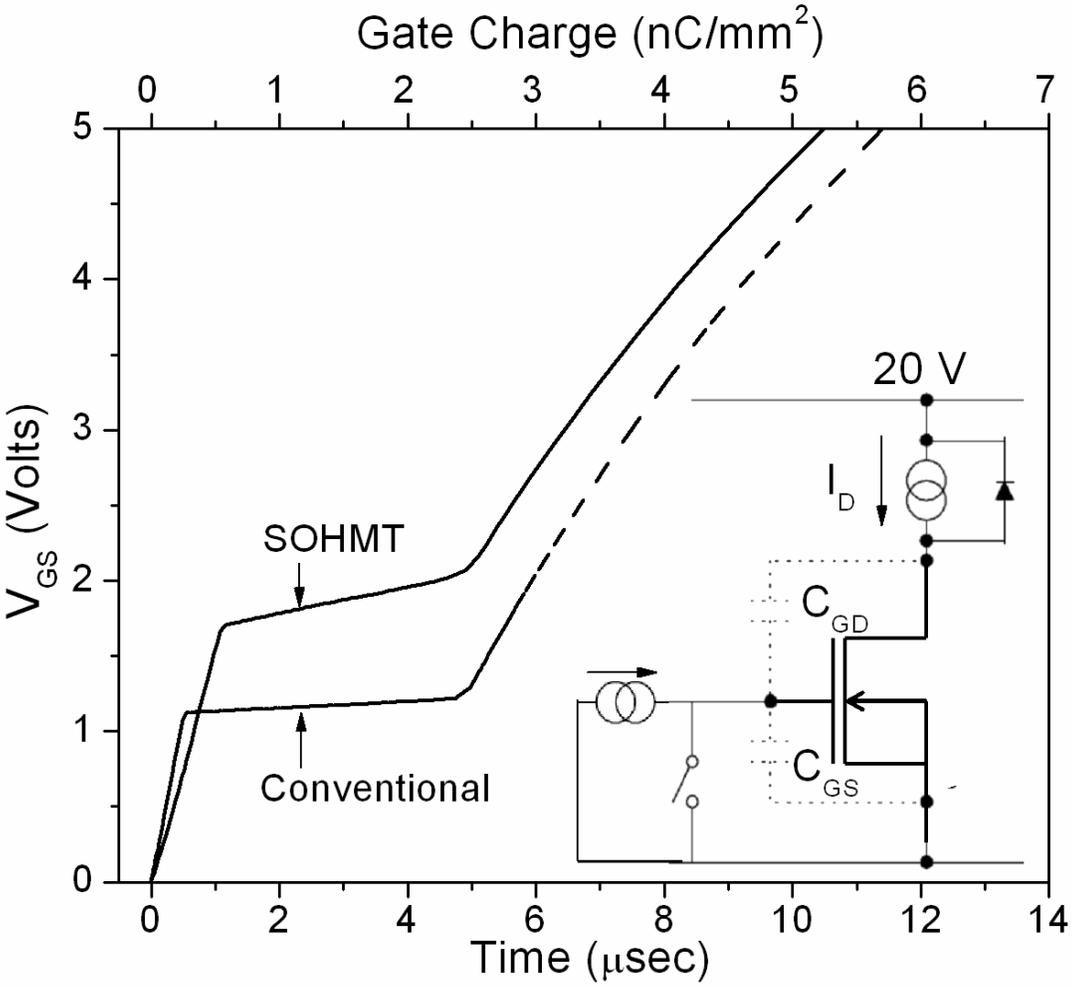

Fig. 5

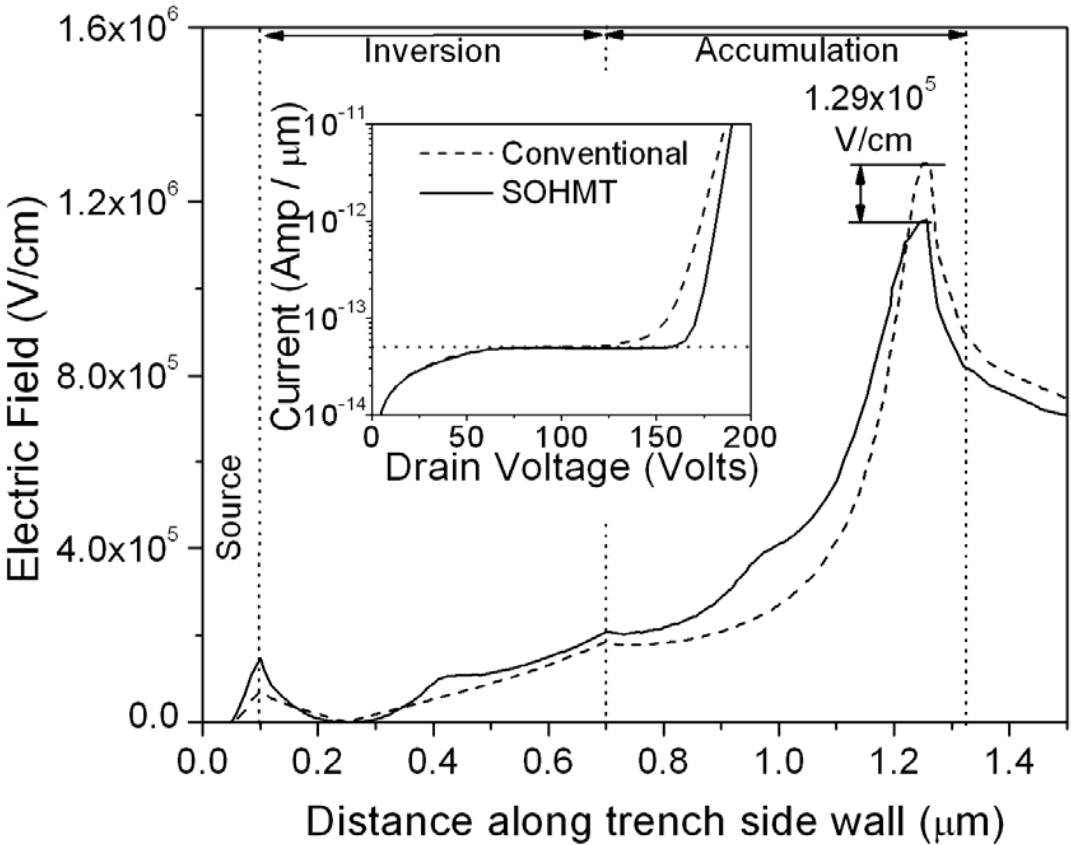

Fig. 6

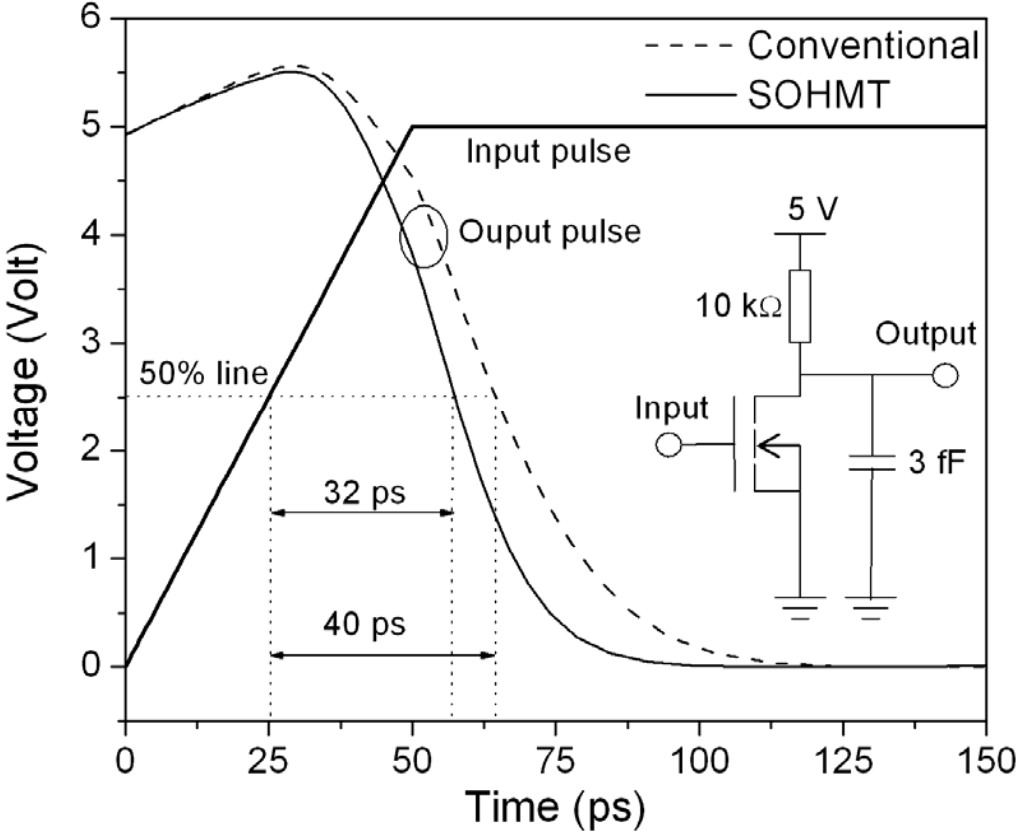

Fig. 7

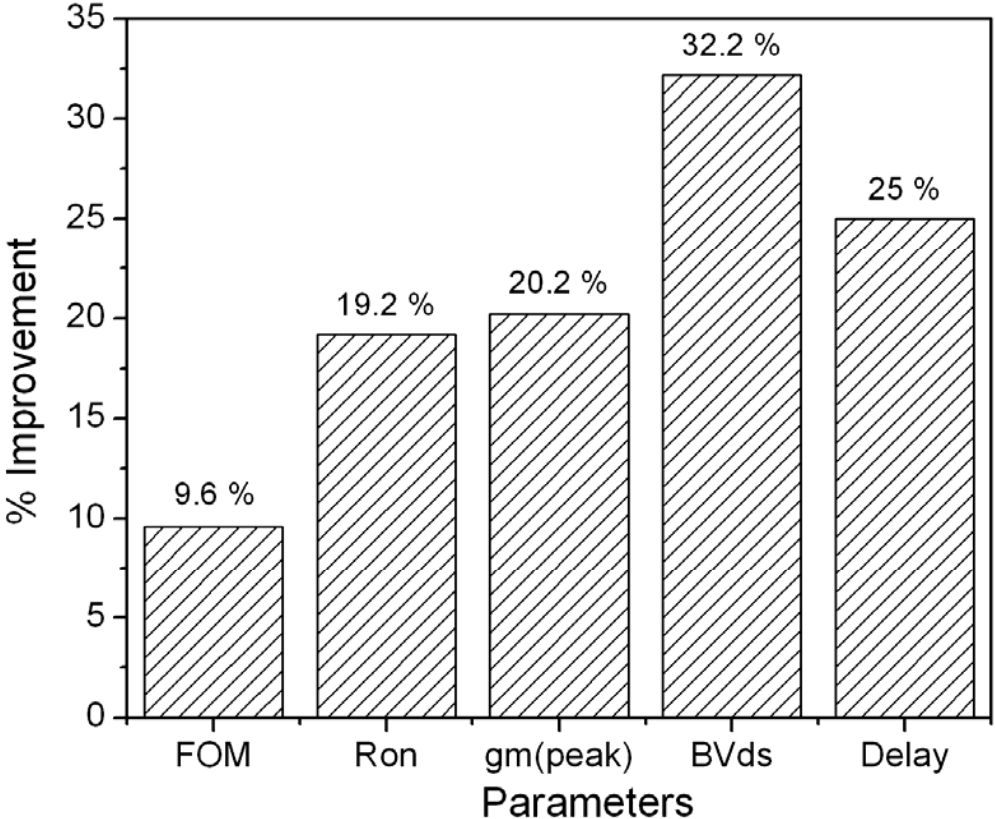

Fig. 8

Raghavendra S. Saxena and M. Jagadesh Kumar "A Stepped Oxide Hetero-Material Gate Trench Power MOSFET for Improved Performance," *IEEE Trans. on Electron Devices,* Vol.56, pp.1355-1359, June **2009.**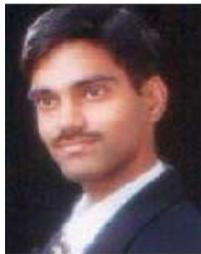

**Raghvendra S. Saxena** received the B.E. degree in electronics and communication engineering from G. B. Pant Engineering College, Pauri Garhwal, India, in 1997, and the M.Tech. degree in microelectronics from the Indian Institute of Technology, Bombay, India, in 2003. He is currently working toward the Ph.D. degree in the Department of Electrical Engineering, Indian Institute of Technology, New Delhi, India.

Since 1998, he has been a Scientist with the Solid State Physics Laboratory (SSPL), Delhi, India, working on the design, modeling, and characterization of infrared detectors and their readout circuits. His current fields of interest are power electronic devices, nanoscale VLSI devices, and infrared detectors. He has published about twenty papers in various international refereed journals and conference proceedings in the above fields. He has also been a reviewer for the journals *IEEE Electron Device Letters*, *IEEE Sensors Journal* and *Recent Patents in Electrical Engineering*.

Mr. Saxena is a corporate member of the Institution of Electronics and Telecommunication Engineers (IETE), India.

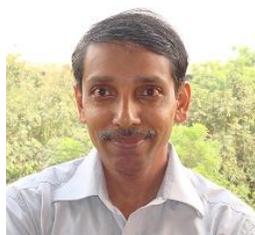

**M. Jagadesh Kumar** (M'95-SM'99) was born in Mamidala, Andhra Pradesh, India. He received the M.S. and Ph.D. degrees in electrical engineering from the Indian Institute of Technology, Madras, India.

From 1991 to 1994, he performed postdoctoral research in modeling and processing of high-speed bipolar transistors with the Department of Electrical and Computer Engineering, University of Waterloo, Waterloo, ON, Canada. While with the University of Waterloo, he also did research on amorphous silicon TFTs. From July 1994 to December 1995, he was initially with the Department of Electronics and Electrical Communication Engineering, Indian Institute of Technology, Kharagpur, India, and then joined the Department of Electrical Engineering, Indian Institute of Technology, Delhi, India, where he became an Associate Professor in July 1997 and a Full Professor in January 2005. His research interests are on Nanoelectronic devices, modeling and simulation for nanoscale applications, integrated-circuit technology, and power semiconductor devices. He has published extensively in the above areas with three book chapters and more than 130 publications in refereed journals and conferences. His teaching has often been rated as outstanding by the Faculty Appraisal Committee, IIT Delhi.

Dr. Kumar is a Fellow of (i) Indian National Academy of Engineering (INAE) and (ii) the Institution of Electronics and Telecommunication Engineers (IETE), India. He was a recipient of the 29th IETE Ram Lal Wadhwa Gold Medal for distinguished contribution in the field of semiconductor device design and modeling. He was also the first recipient of ***ISA-VSI TechnoMentor Award*** given by the India Semiconductor Association to recognize a distinguished Indian academician for playing a significant role as a mentor and researcher. He is a recipient of 2008 IBM Faculty Award.

He is an Editor of the IEEE TRANSACTIONS ON ELECTRON DEVICES and Editor-in-Chief of *IETE Technical Review*. He is also on the editorial board of *Journal of Computational Electronics; Recent Patents on Nanotechnology*; *Recent Patents on Electrical Engineering; Journal of Low Power Electronics*; and *Journal of Nanoscience and Nanotechnology*; He has reviewed extensively for different international journals.

Dr. Kumar is an IEEE Distinguished Lecturer of Electron Devices Society. He is also a member of the EDS Publications Committee and EDS Educational Activities Committee.

21